\documentclass[conference]{IEEEtran}
\IEEEoverridecommandlockouts
\usepackage{cite}
\usepackage{amsmath,amssymb,amsfonts}
\usepackage{algorithmic}
\usepackage{graphicx}
\usepackage{textcomp}
\usepackage{xcolor}
\def\BibTeX{{\rm B\kern-.05em{\sc i\kern-.025em b}\kern-.08em
    T\kern-.1667em\lower.7ex\hbox{E}\kern-.125emX}}

\usepackage[printwatermark]{xwatermark}
\usepackage{soul}
\usepackage{xspace}
\usepackage{multirow}
\usepackage{fixltx2e}
\usepackage{soul}% for underlines

\usepackage{subcaption}
\usepackage{listings}
\usepackage{balance}
\usepackage{enumitem}
\usepackage{graphicx}
\usepackage{booktabs}
\definecolor{mGreen}{rgb}{0,0.6,0}
\definecolor{mGray}{rgb}{0.5,0.5,0.5}
\definecolor{mPurple}{rgb}{0.58,0,0.82}
\definecolor{backgroundColour}{rgb}{0.95,0.95,0.92}

\usepackage[framemethod=tikz]{mdframed}
\usetikzlibrary{shadows}
\usepackage{graphics}
\newmdenv[
    tikzsetting= {fill=blueish!15},
    skipabove=0.33em,
    skipbelow=0.33em,
    linewidth=1pt,
    innerleftmargin=4pt,
    innerrightmargin=4pt,
    innertopmargin=2pt,
    innerbottommargin=2pt,
    linecolor=gray95,
    roundcorner=2pt, 
    shadow=true,
    shadowsize=4pt,
    shadowcolor=black
]{myshadowbox}
\usepackage{tikz}

\lstdefinestyle{CStyle}{
    backgroundcolor=\color{backgroundColour},   
    commentstyle=\color{mGreen},
    keywordstyle=\color{magenta},
    numberstyle=\tiny\color{mGray},
    stringstyle=\color{mPurple},
    basicstyle=\footnotesize,
    breakatwhitespace=false,         
    breaklines=true,                 
    captionpos=b,                    
    keepspaces=true,                 
    numbers=left,                    
    numbersep=5pt,                  
    showspaces=false,                
    showstringspaces=false,
    showtabs=false,                  
    tabsize=2,
    language=C,
    belowskip=0mm,
    moredelim=[is][\color{red}\underbar]{@}{@}
}

\lstdefinestyle{customc}{
  belowcaptionskip=1\baselineskip,
  breaklines=true,
  frame=L,
  xleftmargin=\parindent,
  language=C,
  showstringspaces=false,
  basicstyle=\footnotesize\ttfamily,
  keywordstyle=\bfseries\color{green!40!black},
  commentstyle=\itshape\color{purple!40!black},
  identifierstyle=\color{blue},
  stringstyle=\color{orange},
  moredelim=[is][\color{red}]{@}{@}
}

\lstdefinestyle{customasm}{
  belowcaptionskip=1\baselineskip,
  frame=L,
  xleftmargin=\parindent,
  language=[x86masm]Assembler,
  basicstyle=\footnotesize\ttfamily,
  commentstyle=\itshape\color{purple!40!black},
}

\newcommand{\etal}{\hbox{\emph{et al.}}\xspace}
\newcommand{\eg}{\hbox{\emph{e.g.,}}\xspace}
\newcommand{\ie}{\hbox{\emph{i.e.,}}\xspace}

\newcommand{\etc}{\hbox{\emph{etc.}}\xspace}
\newcommand{\juliet}{\hbox{\sc Juliet}\xspace}
\newcommand{\julietbf}{\hbox{\sc \textbf{Juliet}}\xspace}
\newcommand{\dataset}{\hbox{\sc D2A}\xspace}
\newcommand{\datasetbf}{\hbox{\sc \textbf{D2A}}\xspace}

\newcommand{\tool}{{\sc Velvet}\xspace}
\newcommand{\tools}{{\sc Velvet}'s\xspace}
\newcommand{\toolensem}{{\sc Velvet-ensemble}\xspace}
\newcommand{\toolggnn}{{\sc Velvet-ggnn}\xspace}
\newcommand{\tooltrans}{{\sc Velvet-transformer}\xspace}
\newcommand{\toolbf}{\hbox{\sc \textbf{Velvet}}\xspace}
\sethlcolor{lightgray}
\definecolor{pink}{RGB}{252,145,149}
\definecolor{lightpink}{RGB}{252,145,149}
\definecolor{lightgray}{gray}{0.8}
\definecolor{darkgray}{gray}{0.6}
\definecolor{Gray}{rgb}{0.88,1,1}
\definecolor{Gray}{gray}{0.85}
\definecolor{Blue}{RGB}{0,29,193}
\definecolor{MyDarkBlue}{rgb}{0,0.08,0.45} 
\definecolor{pink}{RGB}{231,95,110}
\definecolor{greenish}{RGB}{182, 231, 142}
\definecolor{orangish}{RGB}{255, 206, 144}
\definecolor{lavender}{RGB}{225, 213, 231}
\definecolor{lightergray}{rgb}{0.85, 0.85, 0.85}
\definecolor{lightestgray}{rgb}{0.95, 0.95, 0.95}
\definecolor{codebg}{HTML}{F4F4F4}
\definecolor{blueish}{RGB}{177, 206, 232}
\definecolor{gray05}{gray}{0.95}
\definecolor{gray10}{gray}{0.90}
\definecolor{gray15}{gray}{0.85}
\definecolor{gray20}{gray}{0.80}
\definecolor{gray25}{gray}{0.75}
\definecolor{gray30}{gray}{0.70}
\definecolor{gray35}{gray}{0.65}
\definecolor{gray40}{gray}{0.60}
\definecolor{gray45}{gray}{0.55}
\definecolor{gray50}{gray}{0.50}
\definecolor{gray55}{gray}{0.45}
\definecolor{gray60}{gray}{0.40}
\definecolor{gray65}{gray}{0.35}
\definecolor{gray70}{gray}{0.30}
\definecolor{gray75}{gray}{0.25}
\definecolor{gray80}{gray}{0.20}
\definecolor{gray85}{gray}{0.15}
\definecolor{gray90}{gray}{0.10}
\definecolor{gray95}{gray}{0.05}

\begin{document}

\title{VELVET: a noVel Ensemble Learning approach to automatically locate VulnErable sTatements
}

\author{
\IEEEauthorblockN{Yangruibo Ding\IEEEauthorrefmark{1}, Sahil Suneja\IEEEauthorrefmark{2}, Yunhui Zheng\IEEEauthorrefmark{2}, Jim Laredo\IEEEauthorrefmark{2}, Alessandro Morari\IEEEauthorrefmark{2}, Gail Kaiser\IEEEauthorrefmark{1}, Baishakhi Ray\IEEEauthorrefmark{1}}
\IEEEauthorblockA{\IEEEauthorrefmark{1}Columbia University, \IEEEauthorrefmark{2}IBM Research}}
\maketitle

\thispagestyle{plain}
\pagestyle{plain}

\begin{abstract}
Automatically locating vulnerable statements in source code is crucial to assure software security and alleviate developers' debugging efforts. This becomes even more important in today's software ecosystem, where vulnerable code can flow easily and unwittingly within and across software repositories like GitHub.  Across such millions of lines of code, traditional static and dynamic approaches struggle to scale. Although existing machine-learning-based approaches look promising in such a setting, most work detects vulnerable code at a higher granularity -- at the method or file level. Thus, developers still need to inspect a significant amount of code to locate the vulnerable statement(s) that need to be fixed. 

This paper presents \toolbf, a novel \textit{ensemble learning} approach to locate vulnerable statements. Our model combines graph-based and sequence-based neural networks to successfully capture the local and global context of a program graph and effectively understand code semantics and vulnerable patterns. 
%From our ablation study, we empirically reveal that the graph-based model is effective in capturing local data dependencies while the sequence-based model focuses more on long-range dependencies. 
To study \toolbf's effectiveness, we use an off-the-shelf synthetic dataset and a recently published real-world dataset. In the static analysis setting, where vulnerable functions are not detected in advance, \toolbf achieves 4.5$\times$ better performance than the baseline static analyzers on the real-world data. For the isolated vulnerability localization task, where we assume the vulnerability of a function is known while the specific vulnerable statement is unknown, we compare \toolbf with several neural networks that also attend to local and global context of code. \toolbf achieves 99.6\% and 43.6\% top-1 accuracy over synthetic data and real-world data, respectively, outperforming the baseline deep learning models by 5.3-29.0\%. 
%\gail{need to clarify that there are two different baselines for the two different tasks.  the first is stated as static analyzers, the second needs to be stated as deep learning}

%This paper presents \tool, a novel \textit{ensemble learning} approach to automatically locate the vulnerable statements. Our model combines the graph-based and sequential-based neural networks, which successfully captures the local and global context of a program graph to effectively understand code semantics and vulnerable patterns. To study \tool's effectiveness, we use an off-the-shelf synthetic dataset and a recently published real-world dataset. \textcolor{red}{For the isolated vulnerability localization task}, \tool achieves 99.6\% and 43.6\% top-1 accuracy over synthetic data and real-world data respectively, outperforming the next-best related work we studied by 5.3\%. \textcolor{red}{In the static analysis setting, where vulnerability is not detected in advance}, \tool achieves 4.5$\times$ better performance than the baseline static analyzers on the real-world data.

%\gail{what are the two tasks/settings? the reviewer will not understand what is the difference between isolated vulnerability localization and vulnerability not detected in advance}

%\gail{was only top-1 considered?  why not top-3, etc.?  does the paper explain why top-1 was chosen, e.g., as a balancing point for false positives / false negatives? Could n for top-n be configured or is 1 hardwired?}

%\gail{shouldn't the ccs concepts include something about security and vulnerabilities?}
\end{abstract}

\begin{IEEEkeywords}
Security Bugs, Vulnerability Localization, Ensemble Learning, Transformer Model, Graph Neural Network
\end{IEEEkeywords}

\vspace{-3mm}
\section{Introduction}\label{sec:intro}

Rapid detection and elimination of vulnerabilities is crucial to protect production software from malicious attacks.  Unfortunately, the shortcomings of traditional program analysis and software testing techniques become apparent at the scale of the software nowadays~\cite{johnson2013don, smith2015questions,liu2012software}. 
%\gail{most of these papers are too old.  the problems in 2005 are not today's problems, cloud computing and devops didn't exist yet. the problems of 2015 might not even be today's problems, can you replace all these with cites from 2018 or later? the same goes for most of the motivation, background and related work cites throughout this paper.}
For example, dynamic analysis tools are known to suffer from high false negatives, as they cannot reach many code regions, particularly given the huge size of modern applications and infrastructure. Static analysis tools scale better but require configuration with known vulnerability patterns (i.e., rules), typically running behind the attackers, and tend to report high false positives. 
%\gail{can tool detect previously unknown vulnerabilities? presumably not, I'd guess it can only detect the kinds of vulnerabilities its trained on, by definition already well-known by the time someone puts together the training dataset, but detecting zero day is the holy grail}

Recent progress in AI techniques, combined with the availability of large volumes of source code, presents an opportunity for security analysts to apply data-driven approaches that augment traditional program analysis. Researchers have explored applying deep-learning techniques to identify security vulnerabilities~\cite{suneja2020learning, buratti2020exploring,russell2018automated, li2018vuldeepecker, li2018sysevr, li2017large, maiorca2019digital, suarez2017droidsieve, Zhou2019DevignEV, chakraborty2020deep, xiao2021deepwukong}. These works typically learn vulnerability patterns from large amounts of vulnerable/non-vulnerable examples without active manual effort. However, previous approaches are mostly limited to predicting vulnerable methods or files, without \emph{locating} the statement that really triggers the vulnerability. Such coarse-grained vulnerability detection slows down developers seeking to locate and fix a vulnerability, since they still need to spend significant debugging effort to inspect hundreds or even thousands of lines of source code manually.  
%\gail{does tool only work on source code, or could it work on bytecodes/binaries?}

However, it is challenging to locate vulnerabilities at the finer granularity of identifying vulnerable statements.  
%\textcolor{purple}{From a machine learning perspective, localization vs. classification is analogous to asking a visual recognition model to not only identify a cat image but to localize exactly which pixels were key to representing the cat.}
%\gail{the third reviewer for fse research track is unhappy about the analogy to visual recognition, which is indeed very weak, individual pixels in images is not akin to individual statements in methods, pixels seem more like individual characters in the source code. in any case, the localization vs. classification problem is phrased badly, since someone could classify each individual statement as vulnerable vs. non-vulnerable, making it indeed a classification problem equivalent to localization.  the challenge needs to be phrased in terms of finer granularity, not in terms of classification vs. localization}
%Vulnerability localization also comes with some practical concerns:  
First, existing vulnerability detection tools~\cite{Zhou2019DevignEV, russell2018automated, suneja2020learning} classify the function as a whole, and a recent research study~\cite{chakraborty2020deep} revealed that these tools learn high-level vulnerable features and cannot highlight the individual vulnerable statements. In contrast, localization requires the model to learn more concrete statement-level vulnerable features; the model needs to pay attention not only to the individual 
statements but also to the control flows and data dependencies among them.
%\gail{Can you add something about why 'paying attention' to context is hard? to classify vulnerable methods, didn't those ML models need to pay attention to the contexts of the methods, e.g., the other methods called in those methods, the other methods that provide the parameters, etc.}
Second, manually annotating vulnerable statements requires significant effort, so collecting a large volume of reliable training data containing vulnerable location information is expensive.
%\gail{do you have any volume, large or otherwise, of *manually* annotated vulnerable statements?  seems like something close to this manual annotation could be drawn from previous fixes to vulnerabilities - the specific statements that were edited. a vulnerability localization tool needs to find the statements for developers to edit, not some other statements that "contain the vulnerability" if the developers still have to search a lot of code to find the specific statements to edit.}
%\gail{Separate from that “The Importance of Accounting for Real-World Labelling When Predicting Software Vulnerabilities” from Jimenez et al., mentioned by the third reviewer of the tools track, should be cited but I do not see it on cite list}
%\rdc{I actually do not agree with this reviewer, since localization is our main task, and in the granularity of statements, the dataset is actually super imbalanced}
We address these challenges by 
(i) developing a novel \textit{ensemble learning} approach, \tool, that learns to capture code semantics at statement granularity from both local and global context. 
(ii) pre-training on large amounts of synthetic data to learn artificial vulnerability patterns, and then fine-tuning on a smaller real-world dataset, which enables the model to understand more complex patterns even though large real-world annotated datasets are not available.

\noindent
\textbf{Modeling Vulnerability Localization.}
We propose \tool to locate vulnerable statements. Our design stems from two insights: (i) the model needs to capture the semantics of the vulnerable statements, and (ii) the semantics often depend on both local and global context. To this end, \tool consists of two main steps:

\noindent
\textit{(i)~Learning Node Semantics.} For locating a vulnerable statement, it is important to understand the statement semantics (\eg control and data dependency, context, \etc). In a static analysis setting, such semantics can be captured well with a code graph, where each graph node represents code elements and edges represent the dependencies between the nodes. Representing these dependencies via a graph has proven effective to understand the code syntax and semantics by many previous studies~\cite{allamanis2018learning, yin2018learning, Hellendoorn2020Global, Zhou2019DevignEV, Dinella2020HOPPITY, chakraborty2020deep, dash2018refinym, suneja2020learning, wang2020ginn, gao2019binseeker}. In this work, we use a Code Property Graph (CPG)~\cite{yamaguchi2014modeling} to represent the code. We then use a Deep Learning (DL) model to learn node semantics from the CPG. The DL model essentially learns a node-level embedding that captures the node content along with contextual dependencies. We then feed the embedded node representation to a classifier to classify the node as vulnerable or not. We further map the node back to the corresponding statement as the final prediction.

\noindent
\textit{(ii)~Incorporating Global and Local Context with Ensemble Learning.} To locate the vulnerable statement, developers spend significant amounts of time debugging the program, trying to both understand the functionality at a high level and, also, estimate the behaviors from local context of suspicious code blocks. Listing~\ref{list:motivate} shows a confirmed CVE from our real-world dataset. To identify the out-of-array access at line 146, developers need to know the latest update of \texttt{block\_ptr}, which is in the surrounding context (line 138), together with the faraway declarations of variable \texttt{row\_ptr} and \texttt{pixel\_ptr}. Thus, to locate the vulnerable line, developers need to reason about both \emph{local} and \emph{global} context. To mimic real-world debugging practice, we propose a novel ensemble approach to learn both contexts for vulnerability localization.

Using our \tool framework, we capture node semantics with a transformer-based model and a GGNN model. 
We use a linear layer and a softmax to assign vulnerability probabilities to each embedded node.
The node with the highest vulnerability probability will be marked as vulnerable, and will be mapped back to the source code statement. Our ablation study shows that \textit{transformer is better able to capture long-range dependencies, i.e., global context, while GGNN captures short-range local dependencies better}. Thus, we combine these two models as an ensemble for final localization.

\noindent\textbf{Pre-training \& Fine-tuning.} 
Another challenge for DL-based vulnerability localization tools is the scarcity of  high-quality, real-world vulnerability datasets with well-annotated statement-level information, \ie which specific line or lines are the triggers of the security issue (\eg, line 146 in Listing~\ref{list:motivate}). Fortunately, a recently launched real-world vulnerability dataset, \dataset~\cite{D2A}, contains precise vulnerability location information. We apply \dataset as the main resource on which to build the prototype for our data-driven technique. To imitate the practical scenario at best, we sort the functions in \dataset based on their commit dates. We train the model on the \emph{past} commits and evaluate the performance on the \emph{latest} ones.

%\gail{what is the difference between root cause and trigger? not clear which you intend to refer to here}

%\gail{One of the reviewers wants more info about D2A.  There's a bit in the response which could be placed in the body of the paper. Should also mention in the paper that the 5\% held out for testing is the most recent data.}

Because of the time-consuming data collection process and the expensive manual validation, \dataset has a relatively small size and is not enough to sufficiently train a generalized model. One possible alternative is to train a localization model on synthetic vulnerabilities: for example, NIST's Juliet Test Suite~\cite{nist2018juliet} was artificially produced to imitate CWE~\cite{mitre2020cwe} vulnerability patterns, and it indicates the location information for each sample. However, models trained on synthetic data do not usually perform very well in real scenarios~\cite{chakraborty2020deep}. In this work, we propose a practical mitigation to address these concerns: we pre-train \tool with large-scale synthetic data, forcing the model to first learn simple, artificial patterns, and then fine-tune with \dataset examples to capture the complexity of real-world patterns. We show that the pre-training and fine-tuning workflow mitigates the dataset scarcity problem.

%\vspace{-3mm}

\noindent
\textbf{Results.}
In the static analysis setting, where vulnerable functions are not detected  in advance, \tool achieves 2.7$\times$ and 4.5$\times$ better performance than baseline static analyzers on the synthetic and real-world dataset separately, and significantly reduces the false positives and false negatives. For the vulnerability localization task, where we assume the vulnerable function has already been perfectly detected, \tool achieves 99.6\% accuracy on the NIST Juliet Test Suite~\cite{nist2018juliet}. With further fine-tuning on the \dataset dataset, \tool achieves 43.6/63.9\% localization accuracies for top-1/3 predictions over the much more complex real-world data. Our ablation study further provides evidence that Transformer is better at capturing distant dependencies while GGNN focuses better on local context of statements, and our \tools ensemble approach is the most effective way of combining global and local information for vulnerability localization -- outperforming the baseline deep-learning models by 5.3\%-29.0\%. %Our datasets and replication package are publicly available here: \url{https://github.com/Anonymous-authors-2020/VELVET}.

\vspace{-1mm}
\begin{lstlisting}[style=customc, belowcaptionskip=0pt, caption= \small Out-of-array accesses vulnerability: CVE-2013-7009, label={list:motivate}, escapechar=\%]
// project: ffmpeg (commit sha: 920046a) 
// file: libavcodec/rpza.c
1  static void rpza_decode_stream(...) 
2  {
14 unsigned short *pixels = ...;
15 int row_ptr = 0;
16 int pixel_ptr = 0; // Fix: int pixel_ptr = -4;
   ...
135  case 0x00:
136    if (s->size - stream_ptr < 16)
137      return;
138    block_ptr = row_ptr + pixel_ptr;
   ...
146    @pixels[block_ptr] = colorA;@
147    block_ptr ++;
148    }
149    block_ptr += row_inc;
162}
\end{lstlisting}
\vspace{-1mm}

This paper makes the following contributions:

\begin{itemize}[leftmargin=*]
 
    \item We propose an ensemble neural architecture, \tool, to locate vulnerabilities at statement-level by successfully capturing local and global program dependencies~\cite{github_repo}.

    \item We design the model learning process as pre-training on the synthetic data, \juliet, and fine-tuning on real-world data, \dataset, following a practical workflow that alleviates data inadequacy concerns regarding real-world vulnerabilities.

    \item We evaluate \tool on both \juliet and \dataset. Our ablation studies show our ensemble approach significantly reduces false positives and false negatives and performs 4.5$\times$ better than the baseline static analyzers on real-world data. We also show \tools ensemble approach is the most effective way to capture vulnerable contexts among baseline deep-learning models.
    
    %\gail{why are you comparing only to static analyzers rather than also to other DL systems?}
    
%\vspace{-1.5mm}
\end{itemize}

\section{Background}
\label{sec:background}

\noindent\textbf{Graph Neural Network (GNN)}.
A graph is a common way to represent source code, where each code element is modeled as a graph node, and relations between the code elements are captured by the edges. Graph Neural Network (GNN) is a deep-learning model that learns directly from graph structure. GNN has been applied to bug detection~\cite{Zhou2019DevignEV, chakraborty2020deep, li2021vuln} and fixing~\cite{Hellendoorn2020Global, Dinella2020HOPPITY, tarlow2019learning, wang2020ginn}. GNN learns the node representations by aggregating the information through the graph structure where nodes can only communicate via neighboring edges. 
Thus, after sufficient training, each node gains knowledge about its {\em local} neighborhood. In this way, GNN leans more fine-granular information about semantic (\eg, data-dependencies) and syntactic (\eg, tree structure) properties of code than mere token-based representations~\cite{allamanis2018learning}. In this work, we implement GGNN~\cite{li2017gated} as representative of graph-based models.

%\vspace{-1mm}
\noindent\textbf{Transformer Model.}
Transformer model~\cite{vaswani2017} is the state-of-art model for sequence learning and has proven to be effective for source code modeling tasks~\cite{Hellendoorn2020Global, wasi2020transformer, ahmad2020summarization, lutellier2020coconut, ding2020patching, feng-etal-2020-codebert}. Transformer leverages the power of attention mechanisms (\ie self-attention and encoder-decoder attention) and builds the architecture entirely on that basis. Multi-head self-attention enables the model to attend to any set of code elements across arbitrarily long distances so that each element will maintain a \emph{global}~\cite{Hellendoorn2020Global} view over the entire sequence. Thus, compared to GNN, Transformer can aggregate information from two faraway nodes more efficiently. Nevertheless, the Transformer model usually treats code as a sequence of tokens (keywords, variable, \etc) and ignores the underlying graph structure of programs. Theoretically, the multi-head attention is able to capture the relations between code tokens during training, but in practice, learning ``edges", \emph{de novo}, is more challenging than explicitly defining them with graph structure in advance. 

\noindent\textbf{Ensemble Learning.}
Various neural networks (\eg GNN and Transformer) are designed out of divergent instincts and thus have divergent architectures. With the random initialization of model weights, different neural architectures can learn quite distinct aspects of the same dataset, even for the same task. This variance leaves challenges for a single model to capture the complexity of the data distributions. Ensemble learning provides a practical solution to minimize individual architectures' noisy bias and improve overall performance: it aggregates multiple models' decisions by either asking models to vote or (weighted) averaging each model's predictions. Such a combination can build a more stable neural network with a comprehensive understanding of the whole data. The ensemble method has shown its effectiveness in code modeling tasks, \eg Lutellier~\etal~\cite{lutellier2020coconut} use ensemble learning to repair bugs and report better results than a single-model counterpart.

Due to the diversity of vulnerability patterns, one single model will struggle to generalize when locating vulnerabilities. Instead, combining the knowledge learned by multiple neural architectures can be more effective in identifying diverse patterns. To this end, we propose an ensemble approach that integrates the individual advances of GNN and Transformer to predict the vulnerable statements.
%All these three models are proven to be effective in different code modeling tasks but have individual pros and cons because of their distinct intuitions. In this paper, we will evaluate these vanilla models under \tool's framework for vulnerability prediction tasks. 
%To this end, we will study the following 

%\input{p_dataset}
%\vspace{-2mm}
\section{Approach}
\label{sec:approach}

This section presents our \tool framework, which aims to predict the vulnerable statements in source code. 
\tool is built on the premise that similar vulnerabilities have occurred in the past~\cite{kovalenko2018mining, Juergens2009codeclone} and can be learned from the previous experience. In particular, \tool analyzes the code as graphs and tries to identify vulnerable graph nodes. 
%For example, consider the code taken from FFmpeg~\cite{ffmpegurl} in Figure~\ref{fig:ffmpeg}. Here, the red box of source code is the vulnerable location, where an unchecked addition between \texttt{s->pos} and \texttt{len} triggers a potential \texttt{Integer Overflow} issue. We first represent the code as a graph structure (partially shown on the right of source code). \tool identifies the corresponding vulnerable node(s) (marked in red) in the code graph. 

To this end, we train \tool on vulnerable/non-vulnerable samples in a supervised learning setting, where graph nodes are annotated with a vulnerability probability\textemdash true vulnerable nodes are annotated with 1 and non-vulnerable nodes are marked with 0 probability. We then design the vulnerability localization as a classification task where each node is assigned a vulnerability probability. The node with the maximum vulnerability probability will be predicted as the vulnerable location.
With sufficient training, we expect the model to learn the patterns of where vulnerabilities are likely to occur, transplant this knowledge to unseen samples with similar characteristics, and make predictions accordingly.
%\rdc{Check this.}
Limited by the availability of annotated datasets, we currently only consider functions with a single vulnerable statement/node. However, it should be straightforward to extend \tool to locate multiple statements, which we plan as future work once we get such datasets.

%\begin{figure}
%    \centering
%    \includegraphics[width=\linewidth]{figures/problem_formulation_new.png}
%    \caption{\small An example of Integer Overflow illustrating \toolbf's approach of vulnerability localization. The red box in source code indicates the ground-truth vulnerable statement, which corresponds to the red vulnerable node in the code graph. The blue node is the declaration of \texttt{len} and the green node is the closest value update of \texttt{len} prior to the vulnerable node.}
%    \label{fig:ffmpeg}
%    \vspace{-6mm}
%\end{figure}

%\vspace{-1mm}
\subsection{\tools Localization Framework}
\label{subsec: prob_def}
%\vspace{-1mm}

Locating the vulnerable node in a graph can be regarded as classifying graph nodes as vulnerable/non-vulnerable. Thus, we define the vulnerability localization problem with respect to a code graph, where we assume a program can be directly transformed to a graph structure (Section \ref{subsec: build_graphs}). We consider a code graph as a set of nodes and edges $\mathcal{G} = (\mathcal{V, E})$, where $\mathcal{V}$ indicates the set of nodes of the graph and $\mathcal{E}$ represents the list of edges connecting the nodes. %with different types.
Given a graph $\mathcal{G}$, the goal for the vulnerability localization task is to locate the vulnerable node $v \in \mathcal{V}$ by predicting its index,
%~\footnote{With the index of the vulnerable node, we are able to assign that node with the ground-truth vulnerability probability as 1, and all the other nodes as 0.}, 
$y \in |\mathcal{V}|$, where $|\mathcal{V}|$ represents the number of nodes in the graph. Therefore, we build the samples in our dataset as $\mathcal{D} = \{\mathcal{G}_i, y_i\}^m$, where $m$ is the size of our dataset.

%Using such a setting, we model the vulnerability localization task as a classification task.  - repeats above
Given a graph $\mathcal{G}$, this involves three main steps: 
\begin{enumerate}[label=(\roman*), leftmargin=*,itemsep=0pt]
    \vspace{-1mm}
    \item Learn the semantic representation of each node of $\mathcal{G}$ in an embedded space;
    \item Use the node-level semantic embedding to assign a vulnerability probability to each node; and
    \item Select the node with highest vulnerable probability as the vulnerable location of $\mathcal{G}$. 
    \vspace{-1mm}
\end{enumerate}
To identify non-vulnerable graphs, we inject a dummy node to every graph. This dummy node acts as a general representation for the entire graph and indicates whether this graph contains vulnerabilities or not. If a sample is non-vulnerable, we mark the dummy node as the ground-truth location.

%\gail{I agree with reviewer 3 from tools track that this bit about dummy root nodes is incomprehensible.  Need better wording.}

We design a node embedding method $\phi$ (Section \ref{subsec: build_graphs}) to vectorize the nodes $v_i$ as $h_i \in \mathbb{R}^d$, where $d$ is the embedding dimension. The set of $h_i$ that contains all the node embeddings of a graph is defined as $\mathcal{H}$. Given a sample $\{\mathcal{G}, y\}$, we 
%first embed the nodes and 
then feed the vectorized graph into a neural-network model to learn the semantic node representation $\mathcal{H'}=F_{model}(\mathcal{H}, \mathcal{E})$. 
%and update the node representations capturing the node semantics. 
%\begin{align}
%    \mathcal{H} &= [h_1, ..., h_{|\mathcal{V}|}] = [\phi(v_1), ..., \phi(v_{|\mathcal{V}|})] \\
%    \mathcal{H'} &= [h_1', ..., h_{|\mathcal{V}|}'] = F_{model}(\mathcal{H}, \mathcal{E})
%\end{align}
%\brc{H is not defined anywhere}

%An optimal or well-trained model $\mathcal{H'}$ is expected to transform the initial node embeddings to the version with sufficient semantic knowledge, \ie the updated representation of each node should well encode the information of the node itself, together with the significant context.
A well-trained model is expected to encode sufficient semantic knowledge of a node, together with its context, so we feed the transformed semantic representation, $h'$ of each node to a classifier that tries to compute the probability of each node being vulnerable. 
%We realize the classification model with a simple linear layer to predict vulnerability probability. 
More formally, the learned $\mathcal{H'} \in \mathbb{R}^{d_h \times |\mathcal{V}|}$ ($d_h$ is the hidden dimension of the model) is fed into a linear layer with trainable weights $\mathbf{W} \in \mathbb{R}^{d_h \times 1}$ and bias $\mathbf{b} \in \mathbb{R}^{1 \times |\mathcal{V}|}$ to get a score for each node and then a softmax layer for probability $\mathbf{P} = softmax(\mathbf{W}^\top\mathcal{H'} + \mathbf{b})$.
%We feed the learned $\mathcal{H'} \in \mathbb{R}^{d_h \times |\mathcal{V}|}$ ($d_h$ is the hidden dimension of the model) into a linear layer with trainable weights $\mathbf{W} \in \mathbb{R}^{d_h \times 1}$ and bias $\mathbf{b} \in \mathbb{R}^{1 \times |\mathcal{V}|}$ to get a score for each node and then a softmax layer for probability.
%\vspace{-5mm}
%\begin{align*}
    %\mathbf{P} = softmax(\mathbf{W}^\top\mathcal{H'} + \mathbf{b})
%\end{align*}
Given the vulnerable probability of each node, we take the index of the most probable node as the prediction $\hat{y} = argmax_{i \in |\mathcal{V}|}(p_i)$, and compute cross-entropy loss comparing with the ground-truth.
%\begin{align*}
    %\hat{y} &= argmax_{i \in |\mathcal{V}|}(p_i) \\
    %loss &= CrossEntropy(y, \mathbf{P})
%\end{align*}

\begin{figure}
    \centering
    \includegraphics[width=\linewidth]{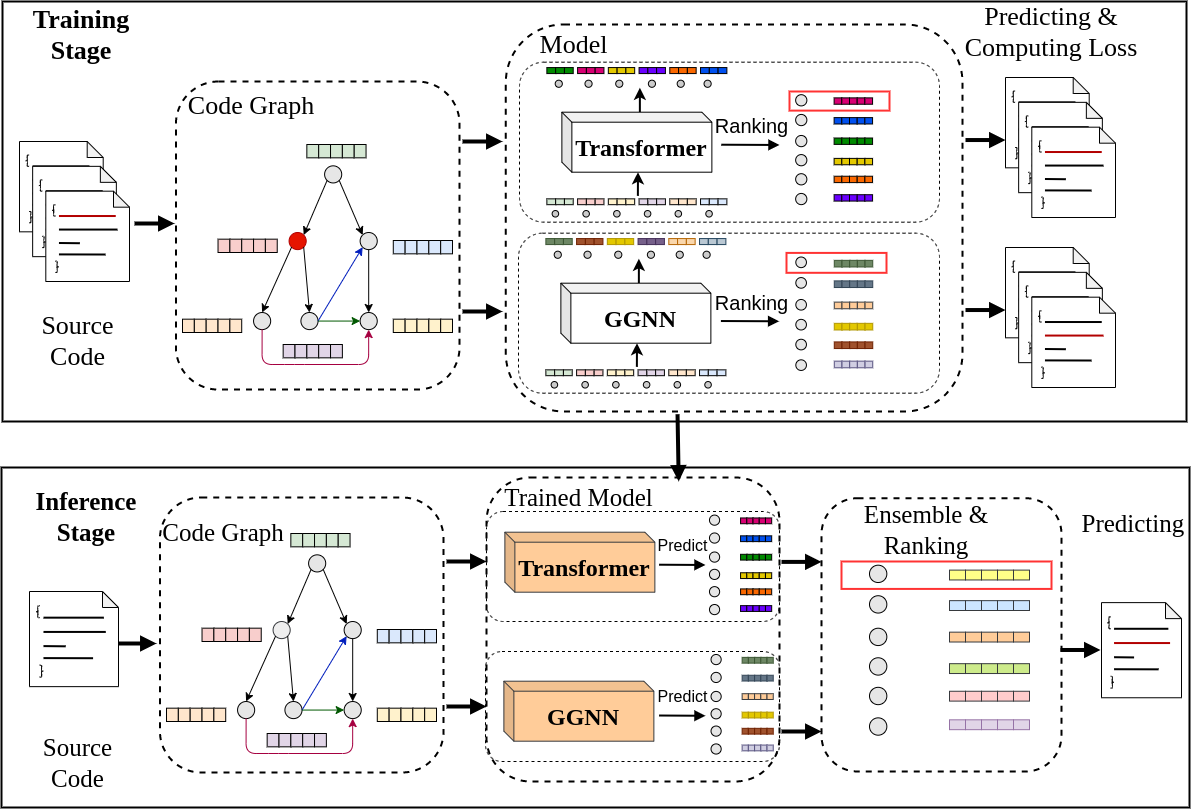}
    \caption{\small \textbf{\toolbf Overview}}
    \label{fig:workflow}
\vspace{-5mm}
\end{figure}

Figure~\ref{fig:workflow} illustrates the end-to-end workflow of the proposed approach. Similar to many NN-based tools, our technique contains two stages: Training and Inference. The 
%We will use the 
following sections 
%to introduce 
explain the details of each module of the framework.
%in details.
%\gail{somehow the third reviewer of fse main track mixed this reference, maybe move to beginning of this section. I checked its still in the same buried location as in that version of the paper}

%\vspace{-2mm}
\subsection{Building Vectorized Code Graphs}
\label{subsec: build_graphs}

\noindent\textbf{Graph Vectorization.}
%To build the input of our model, the first step is to extract the graph structure of code samples. 
In our work, we use Code Property Graph (CPG)~\cite{yamaguchi2014modeling} to represent the graph semantics of programs (code graph module in Fig.~\ref{fig:workflow}).
CPG is a code representation designed specifically for vulnerability detection. Besides AST, CFG, DFG edges, it also includes several other types of edges that encode detailed information regarding program dependencies. 
%We will further evaluate the effectiveness of CPG against other graph structures in Section~\ref{subsec: rq4}. 
Specifically, we use \emph{Joern}~\cite{joerntool} to generate CPG.
%it will output a \emph{nodes.csv} file to indicate the detailed node information including the node type and corresponding code tokens, and also a \emph{edges.csv} file representing edges information (\ie the edge type and the vertices it connects). 
To vectorize the node information, we train a \emph{word2vec} model over all possible code tokens in our dataset, and for each node, we concatenate the token embeddings corresponding to that node, as the initial node representation. 
%As for edges, we consider 10 types of edges, provided by \emph{Joern}, to build CPG, and assign a unique number in the range of 0 to 9 to represent the edge type: \texttt{IS\_CLASS\_OF}, \texttt{IS\_FUNCTION\_OF\_AST}, \texttt{IS\_FUNCTION\_OF\_CFG}, \texttt{IS\_AST\_PARENT}, \texttt{DOM}, \texttt{POST\_DOM}, \texttt{CONTROLS}, \texttt{DECLARES, FLOWS\_TO}, and \texttt{REACHES}.

\noindent\textbf{Vulnerable Node Annotation.} In the graph representation, we annotate the AST node that corresponds to the vulnerable statement as the ground-truth. As shown in Figure~\ref{fig:workflow}, the red line in the training sample represents the vulnerable statement, and we annotate the corresponding (red) node in the graph as the ground truth. 
%Noted that our datasets only include functions with one single vulnerable statement.

%\gail{sounds strange to refer to an AST (tree) as a graph, also noted by the 3rd reviewer}

%\vspace{-2mm}
\subsection{Locating Vulnerable Nodes with Ensemble Model}
%Learning Vulnerable Patterns with Local and Global Model}
\label{subsec: model_learning}

As shown in the example of Listing~\ref{list:motivate}, to identify the integer overflow statement, a localization tool should maintain a global view to understand the initialized value of \texttt{pixel\_ptr} from its declaration, together with the local context about the latest updates of the variable \texttt{block\_ptr}'s value. Capturing such a combination of long-range and short-range dependencies is necessary for the localizer to successfully identify the vulnerable statement. To this end, we propose our ensemble approach built on two main components to capture these two distinct dependencies: GGNN and Transformer.

%\gail{The response says some other models were tried, this could be mentioning briefly in the body of the paper. Is there some reason nothing other than transformers and GGNNs were considered, that is, are there no realistic alternatives for global vs. local?} 

%Researchers have shown that graph-based networks perform well to identify vulnerable patterns in a codebase~\cite{chakraborty2020deep, Zhou2019DevignEV, wang2020ginn}. One of the reasons behind such models' success is that they focus on the semantic order of programs, \ie control-flow, data-flow, \etc, rather than on a left-to-right sequence order. 
%However, 
Graph-based neural networks are effective at understanding the semantic order of programs, since they directly learn control flows and data dependencies with the pre-defined edges. However, training involves a message passing algorithm where nodes only communicate with their neighbors. The ability to learn long-range dependencies is limited by the number of message passing iterations, which are typically set to a small number (\eg less than eight) due to computational cost~\cite{allamanis2018learning, Hellendoorn2020Global, fernandes2018structured, Zhou2019DevignEV, suneja2020learning, chakraborty2020deep}. Such a limitation will result in an inherently \emph{local}~\cite{Hellendoorn2020Global} model. 
%This prevents the model from capturing relations between distant nodes, such as between the first definition of a variable and its last appearance. 
In contrast, Transformer allows \emph{global}, program-wise information aggregation~\cite{Hellendoorn2020Global}, and without pre-defined edges, the self-attention mechanism of Transformer is expected to encode considerable code semantics -- which can be complementary to those defined explicitly by the code graph. Therefore, to learn the diversity of vulnerable patterns, we \textit{separately} train these two distinct models and use their predictions in an \textit{ensemble learning} setting at inference time (Training Stage in Fig.~\ref{fig:workflow}).

%\gail{the second reviewer wants details about how long-range and global the dependencies really are in the datasets - "Are the GNN
%and Transfomer networks restricted to analysing the CPG of a single C++ file, or a single translation unit, or multiple translation units? Many security vulnerabilities involve tracing data-flow across entire translation units (e.g. hundreds of source code files)".  Seems like the paper could mention the longest and average length long-range dependencies that actually occur in the datasets.  The response says "spanning hundreds of lines" but the only example given is 130 lines (146-17+1).  In any case, this example could go in body of paper.  Could also put in body of paper the note in the response about future work on inter-procedure dependencies.}

\noindent\textbf{Local GGNN Model.} As the input, the initial vectorized node representation, $\mathcal{H}$, is passed to the GGNN model, along with the list of edges $\mathcal{E}$. Following the design of Li \etal~\cite{li2017gated}, the GGNN model aggregates information from node-neighbors using  message passing over different edge types with distinct weights. This results in an aggregation vector, representing the information from the neighborhood, for each node at time-step $t-1$. Each node is then, at time-step $t$, updated by aggregating its own representation of $t-1$ and its neighbors' information using GRU units. After K time steps, the output of GGNN model will be the transformed node representations, encoded with \emph{local} context for each node $H^{g} = [h_1^{g}, h_2^{g}, ..., h_{|\mathcal{V}|}^{g}]$
%\begin{align}
%    a_i^{(t-1)} = \sum_e A_e^\top(W_eH'^\top + b)
%\end{align}
%where $e$ is an edge type, $A$ represents the adjacency matrix, $b$ a learnable bias term, and $W$ a learnable weight matrix for each edge type.
%Each node is then updated with its own representation of $t-1$ aggregated from its neighbors using GRU units at time-step $t$. 
%With $h_i$ denoting node representations, this step can be represented as:
%\begin{align}
%    h_i^{(t)} = GRU(h_i^{(t-1)}, a_i^{(t-1)})
%\end{align}
%After certain time steps, the output of GGNN model will be the transformed node representations, .

\noindent\textbf{Global Transformer Model.} In order to learn the implicit code dependencies complementary to GGNN, we only provide Transformer with AST nodes of the graph as a sequence. Removing edges can also enforce the model to learn the long-range dependencies in a more efficient way, since it does not have to reach faraway node step-by-step. We align the AST nodes in the order that reserves the original source code's sequential logic. For example, if the source code has a order of statements as: \texttt{\{stmt1 -> stmt2\}}, then when building the input sequence, all the nodes belonging to \texttt{stmt1} will be placed at the left side of all those belonging to \texttt{stmt2}. 
%We stack six identical transformer encoders, and each encoder has eight attention heads. 
With the power of multi-head self-attention~\cite{vaswani2017}, the final node representations transformed by the model are encoded with a \emph{global} view over the whole function~\cite{Hellendoorn2020Global} $H^{tr} = [h_1^{tr}, h_2^{tr}, ..., h_{|\mathcal{V}|}^{tr}]$

The final node representations from the model (either GGNN or Transformer) are then fed into a multilayer perceptron layer to get a single score for each node: $S^g = [s_1^g, s_2^g, ..., s_{|\mathcal{V}|}^g] \in \mathbb{R}^{|\mathcal{V}|}$ for GGNN, $S^{tr} = [s_1^{tr}, s_2^{tr}, ..., s_{|\mathcal{V}|}^{tr}] \in \mathbb{R}^{|\mathcal{V}|}$ for Transformer. During training, all vulnerability scores are passed into a softmax layer to get the per-node vulnerability probability, and calculate the cross-entropy loss. We note that GGNN and Transformer are trained independently, and each model does its back-propagation and updates weights without interaction. 
%The reason is that we hope to enforce the distinct models to learn diverse aspects of vulnerable patterns at best during Training Stage.

\vspace{-2mm}
\subsection{Ensemble Learning}
\label{subsec: ensemble_learning}

As shown in Figure~\ref{fig:workflow}, in the Training Stagr we independently train two distinct neural architectures, GGNN and Transformer, to force them to learn the diverse aspects of vulnerable patterns. In the Inference Stage, we aim at combining the knowledge of both models to make comprehensive predictions on previously unseen data. To this end, we propose an ensemble approach to aggregate the predictions from both models. Concretely, as shown in Figure \ref{fig:workflow}, we input the vectorized code graph into both well-trained models, and they will output the transformed node representations $H^g$, $H^{tr}$ (Section \ref{subsec: model_learning}) and then the vulnerability scores, $S^g$ and $S^{tr}$ are computed. To aggregate the predictions, we calculate ensemble vulnerability scores, $S^{en}$, for all nodes by averaging: $S^{en} = 0.5 * S^g + 0.5 * S^{tr}$, and then the node with the highest ensemble score will be the predicted vulnerable node. We note that the averaging for ensemble scores can be weighted, but further heuristics need to be introduced to decide the weights for different models. 

The ensemble approach is technically straight-forward, but works quite well in practice. In Section \ref{sec:evaluation}, we show the effectiveness of \tools ensemble and empirically demonstrate that GGNN and Transformer models are indeed learning diverse aspects of vulnerable patterns, as expected intuitively.
%\vspace{-2mm}
\section{Study Design}
\label{sec:design}

In this section, we present our datasets and how we train and test the models with them using distinct evaluation metrics. 

%\vspace{-1mm}
%\vspace{-2mm}
\subsection{Datasets}
\label{sec:dataset}
%\rdc{Explain how we define the ``vulnerable location" here, since such location can be defined as the root cause location or the location that where the vulnerabilities actually happens. In our case, it should be the latter.}

%\vspace{-1mm}
We conducted experiments with the two datasets, {\sc Juliet} (a synthetic dataset) and \dataset (a real-world dataset), which include statement-level vulnerability annotations. Concretely, each sample is a function, and the vulnerable location is the statement that directly exposes the vulnerability (\eg Line 146 in Listing~\ref{list:motivate}). Every sample in the dataset has one single vulnerable statement and is written in the C language.
Note that, there are some other existing C-language vulnerability datasets~\cite{chakraborty2020deep, russell2018automated, nvd, Zhou2019DevignEV}
%(\eg \textsc{ReVeal}~\cite{chakraborty2020deep} and Devign~\cite{Zhou2019DevignEV})
that annotate vulnerabilities only at the function level. Since our goal is to localize vulnerabilities at the statement level, we cannot use them.
%Other datasets, such as VulDeePecker~\cite{li2018vuldeepecker} and SySeVR~\cite{li2018sysevr} (mix of synthetic and real world data) use slices converted into artificial linear sequences. These slices are not valid code statements and they cannot be used to train our model. 

%\vspace{-2mm}
\noindent\textbf{Synthetic Dataset (\julietbf): }
The {\sc Juliet} Test Suite~\cite{nist2018juliet} is a synthetic dataset containing intentional flaws to test static analysis tools. The test cases in the dataset have vulnerabilities covering 118 different Common Weakness Enumeration (CWE) classes~\cite{mitre2020cwe} and well-annotated location information.
%\jal{need reference for CWE} 
%The dataset also provides an xml file that manifests the structure of each test case and indicates which code line or statement is vulnerable, and using this information, 
We extracted around 24,000 vulnerable functions with corresponding faulty locations, and also 26,000 non-vulnerable functions without any vulnerable statements.

Although synthetic datasets provide large amounts of data to train large models ~\cite{li2018vuldeepecker, russell2018automated, li2018sysevr, li2020vuldeelocator}, they fail to capture the complexity of real-world data and tend to restrain the model to simple vulnerable patterns~\cite{chakraborty2020deep}. To overcome this issue, we also apply the recently published \dataset dataset.

\noindent\textbf{Real-world Dataset (\datasetbf): } It is challenging and expensive to collect real-world vulnerabilities with well-annotated statement-level information, so such datasets are rare.
%and thus we can hardly find any off-the-shelf dataset of this kind, with accurate vulnerable locations until recently. - don't know what this means
%To further assist DL-based vulnerability detection and localization tool, IBM researchers published the \dataset~\cite{D2A} dataset. They 
Zheng \etal recently published the \dataset dataset~\cite{D2A} which they collected from multiple C/C++ projects such as {\tt OpenSSL} and {\tt LibTIFF}, which are core components of cloud services. These projects have also been studied by many existing security-related works~\cite{du2019leopard, gao2019binseeker}, but without providing the annotations we need. Zheng \etal collected vulnerabilities using static program analysis and differential analysis.
%~\footnote{Their data collection tool: \url{https://github.com/ibm/D2A}. For further details of this tool, please refer to their publication~\cite{D2A}}. 
Compared to existing datasets, \dataset preserves more details such as function-call traces and locations that trigger the vulnerabilities. The authors also spent much manual effort
to ensure the accuracy of \dataset's labels and location information.

In this work, we derive our real-world dataset from \dataset, and we end up with approximately 2,500 unique functions with annotated vulnerable locations and 2,500 non-vulnerable counterparts in total. Our dataset contains 9 main vulnerability types covering 18 CWE types. Table~\ref{tab:data_distribution} shows the details. The longest function spans 1269 lines of source code, where it would be very challenging for developers to locate the vulnerable statement even if they know the function is problematic. 

Even though both datasets have roughly the same number of vulnerable and non-vulnerable functions, our work focuses on statement-level localization, and at the statement granularity both datasets are actually super \textbf{imbalanced}, aligning well with the relative rareness of vulnerabilities~\cite{chakraborty2020deep, jimenez2019importance}.

\vspace{-2mm}
\begin{table}[h]\centering
\caption{\small 
Distribution of vulnerability types in the datasets.
Column 1\&2 show how the vulnerability types in {\dataset} map to the corresponding CWEs in \juliet. Column 3\&4 show the respective proportions of the type occurrences.
Noted that \juliet covers more vulnerability types than \dataset.
}
\label{tab:data_distribution}
\scriptsize
\begin{tabular}{lrcc}\toprule
\textbf{\datasetbf Type}  &\textbf{{\sc \textbf{\julietbf}}  Type} &\textbf{\sc \textbf{\julietbf}} (\%)  &\textbf{{\sc \textbf{\datasetbf}} (\%)} \\ 
\midrule
Integer Overflow &CWE190-191, 194-197 &21.9\% &56.3\%\\\midrule
Buffer Overrun &CWE121-122, 124-127 &25.0\% &30.8\% \\\midrule
NullPtr Dereference &CWE476 &0.7\% &4.9\%\\\midrule
%Dangling Ptr Dereference &unknown &0.0\% &0.3\%\\\midrule
Memory Leak &CWE401 &2.0\% &0.9\%\\\midrule
Dead Store &CWE563 &1.0\% &1.3\%\\\midrule
Divide by Zero &CWE369 &1.6\% &0.3\%\\\midrule
Null Dereference &CWE690 &1.9\% &3.0\%\\\midrule
Uninitialized Value &CWE457 &1.6\% &1.9\%\\\midrule
Use After Free &CWE416 &0.4\% &0.6\%\\
\bottomrule
\end{tabular}
\vspace{-4mm}
\end{table}

%\vspace{-1mm}
\subsection{Model Pre-Training \& Fine-Tuning}

\vspace{-1mm}
\textit{a) Migration From the Synthetic to the Real-world.} 
As mentioned in Section \ref{sec:dataset}, our real-world dataset, \dataset, is much smaller than \juliet due to the high cost of collecting and annotating the locations of real-world vulnerable statements. 
%This is in part due to high cost of collecting and annotating the locations of real-world vulnerable statements. Collecting large amounts of synthetic data is easier, since the vulnerable patterns are manually designed, and the sample generating process is automated following handcrafted rules. In contrast, gathering real-world vulnerabilities with annotated locations is very expensive. It requires either a conservative and carefully designed heuristic, as in \cite{D2A}, or significant manual effort to validate each candidate vulnerable statement.  Realistically, both approaches can only disclose a rather limited number of examples. 
Given \dataset's small size, the usual same-dataset train-test scheme would likely cause the models to overfit.

% this is not novel, we cannot propose it as a solution
In this work, we adopt a workflow designed to alleviate this concern of data inadequacy by {\em pre-training} \tool on a large amount of synthetic data and {\em fine-tune} it on the limited real-world data. We first sufficiently train models on synthetic data with a relatively large learning rate and then fine-tune the pre-trained models on real-world data with a smaller learning rate. The intuition behind this design is: the ample synthetic data has already exposed a portion of the practical vulnerability distribution, so we ask models to first understand this part with adequate samples, and then keep learning other (more complex) parts when fed with what would otherwise be an inadequate set of real samples. In Section~\ref{subsec: rq4}, we experimentally determine that models successfully leverage this ``pre-training + fine-tuning" setting to migrate the knowledge learned from synthetic data to the real-world scenario.

%\gail{the second reviewer wants more details about D2A: "I would like to see a short description of the range and type of vulnerabilities in the IBM-D2A dataset."  Seems like this should not be hard to add, there's material about this in the response. I also put a note about this in intro.}

%\vspace{-1mm}
\textit{b) Training, Validation, and Testing Split.} 
In the synthetic dataset, \juliet,
%we have \rdc{fix this}24k vulnerable functions with annotated location information, and 
we split the whole dataset into three parts, {\sc train/valid/test}, with a ratio of 90\%:5\%:5\%. We train \tool on the {\sc train} split from scratch for 10 epochs and evaluate the performance on {\sc test} split. 

%In the real-world dataset, \dataset, we have 2.5k vulnerable functions. %\rdc{Need to highlight this in some ways.} 
For \dataset, to imitate the practical scenario at best where we learn from previous vulnerabilities, we sort the functions based on their commit dates. We then \emph{train the model on the past commits and evaluate the performance on the latest ones}. Again, we split the sorted samples into {\sc train/valid/test} with the ratio of 90\%:5\%:5\% ({\sc test} split is made up of the latest 5\%). %\rdc{check this}
%\gail{'latest' may need more emphasis, since a reviewer missed it, maybe something like the bit in response about learning from older data applied to newer data}
We fine-tune the trained models from the synthetic dataset on {\sc train} split for 50 epochs and evaluate on {\sc test} split. We note that the numbers of epochs for training are carefully selected: we ensure that, in such settings, the validation loss on {\sc valid} split of all the models will no longer decrease for 3 consecutive epochs.

%\vspace{-1mm}
\textit{c) Experimental Configuration} Models are configured with the input hidden dimension of 256, a dropout rate of 0.1. The learning rate for pre-training is $10^{-4}$, and for fine-tuning is $10^{-5}$. We use Gated Graph Neural Network (GGNN) model~\cite{li2017gated} with 8 time steps, similarly to several code modeling works~\cite{allamanis2018learning, Dinella2020HOPPITY, Zhou2019DevignEV, Hellendoorn2020Global, chakraborty2020deep, suneja2020learning}. For the Transformer, we use 6 encoder layers, 8 attention heads, while the attention dimension is 512 and feedforward layers of dimension 2048. 
All models are implemented using Tensorflow 2.2.0, CUDA 10.1, and trained using 8GB Nvidia GeForce RTX 2080 SUPER GPU. For data processing, our machine takes 26.8 seconds to generate every 1000 CPGs, and takes 60.9 seconds to vectorize them. It takes 32 minutes to train word2vec embeddings. 
% why is this relevant?
The model training on \juliet takes ~14 hours, and ~24 hours on \dataset. These numbers are at par with existing deep-learning-based vulnerability analysis~\cite{Zhou2019DevignEV, chakraborty2020deep, Hellendoorn2020Global}.

\vspace{-1mm}
\subsection{Evaluation of Statement-level Localization}
\label{sec:metric}

%\rdc{You did not talk about mapping node to statement}

%\noindent\textbf{Line-level evaluation.} 
As we introduced in Section \ref{subsec: prob_def}, the model prediction will be the index of a graph node. To evaluate the localization performance in a practical scenario, we maintain a mapping between a node and its corresponding line number in the source code. We map both the ground-truth node and the predicted node to their source line number and check whether they match. We evaluate statement-level localization performance using the following metrics:

\noindent\textbf{Top-k Accuracy.} 
We use top-k accuracy as one metric to evaluate the localization quality. We map the top-k predicted nodes with the highest scores to their source line numbers, and if at least one of the predictions matches the ground-truth, we define the prediction as correct. %For \juliet evaluation, we only show the results for top-1, since it is a synthetic dataset with simple distributions, the models can already achieve nearly perfect performance for just top-1. For \dataset, we show the results for top-1 and top-3, since we assume $k = 3$ is an acceptable setting in practice.

\noindent\textbf{Prediction Distance.} 
While accuracy is an important metric, it fails to fully capture realistic aspects of the vulnerability localization problem. For instance, when a wrongly predicted location is not too far from the true target, the developer can still find the vulnerability by glancing at the surrounding code. In order to evaluate this aspect, we measure how far the prediction is from the true location. We define the distance between these two as \emph{Prediction Distance} and we calculate it as $Distance = \lvert l_{pred} - l_{true} \rvert$, where {$l_{pred}$} is the line number of the model prediction, and {$l_{true}$} is the ground-truth. 
%The prediction distance will be used as an additional evaluation metric in our coming experiments.

\vspace{-2mm}
\subsection{Baselines}
\label{sec:baselines}

\noindent
\textbf{Static Analyzers}
%\rdc{check this} 
One major motivation of deep-learning-based vulnerability detection tools is to overcome the drawbacks of rule-based static analyzers. We use them as baselines to show the improvement brought by \tool. We pick three open-source static analyzers, Infer~\cite{calcagno2011infer}, FlawFinder~\cite{flawfinder} and RATS~\cite{rats}. These are popular and frequently used by developers and researchers. For example, Infer is widely used by large enterprises~\cite{calcagno2015moving, D2A} and FlawFinder is integrated into many open-source code scanners such as GitHub Code Scanner~\cite{githubcodescanner} and Codacy Security Scan~\cite{codacy}. These three static analyzers are also used as baselines by work on deep-learning-based vulnerability detectors~\cite{li2018vuldeepecker, russell2018automated, li2020vuldeelocator, xiao2021deepwukong}. 
%\textbf{Localization Performance} There are very few tools in the literature that have used DL based techniques to build a static analysis tool for locating vulnerability at statement level.\footnote{We further discuss related work in Section~\ref{sec: related}} Most of the existing tools try to detect vulnerabilities at function levels. The best we can do in this situation is to compare against static analysis tools, which we have done in Section~\ref{subsec: rq2}.

\noindent
\textbf{Comparing Architectures under \toolbf's framework}
To figure out the best neural architecture that can be fit into our framework for vulnerability localization, we compare different models by replacing the architecture in the model module in Figure~\ref{fig:workflow}. 
For example, we compare the ensemble model with GGNN-only and Transformer-only models in Section \ref{subsec: rq2} and \ref{subsec: rq1}, resp. 
Note that the transformer here is a bit different than the commonly used transformer of CodeBert~\cite{feng-etal-2020-codebert}, Roberta~\cite{liu2019roberta} \etc , where the inputs are token sequences. In our case, the input consists of pre-processed graph nodes, as we use node embedding instead of token embedding. %To avoid the confusion, we annotate our ensemble approach as \toolensem and the baselines as \toolggnn and \tooltrans (short for Transformer).

\noindent
\textbf{Comparing Aggregation Strategies under \toolbf's framework} 
\tool aggregates the information learnt from \emph{global} and \emph{local} contexts as an ensemble. Other researchers have tried different aggregation strategies:
% ---although the applications were different. - says same thing below with better wording
Hellendoorn \etal~\cite{Hellendoorn2020Global} propose Graph-Sandwich models, which stack sequence-based model and graph-based model to capture the \emph{global} and \emph{local} semantics for detecting and fixing variable-misuse bugs. They also propose GREAT, which integrates edge information of code graphs into the Transformer model. 
Dinella \etal~\cite{Dinella2020HOPPITY} propose a graph-based model to learn bug-fixing edits, where they incorporate a \emph{global} pointer mechanism to locate the buggy node. 
These previous works share similar insights of aggregating information from diverse contexts, although for different tasks. 
We adapted these methods in our \tools framework to compare different ways to aggregate \emph{global} and \emph{local} information for vulnerability localization. We implement Transformer-sandwich, GREAT, and the localization part of Hoppity by reusing their open-source packages \cite{Hellendoorn2020Global, Dinella2020HOPPITY}.
To compare under identical settings, 
%we adapt these models into the \tool framework---
we replace the model part in Figure \ref{fig:workflow} with the different architectures and keep all the other parts as-is.
%\vspace{-3mm}

%We name these adapted models as 

%\vspace{-2mm}
\section{Results}
\label{sec:evaluation}
\begin{table*}[h]\centering
%\vspace{-2mm}
\caption{\small Results when jointly training models on vulnerable and non-vulnerable functions. As discussed in Section \ref{subsec: rq2}, Pred Acc. is the \emph{Prediction Accuracy}, Vul-CLS indicates the \emph{Vulnerability Classification} setting, and Vul-LOC Acc. is the top-1 localization accuracy.}
\vspace{-1mm}
\label{tab: rq1_full}
\scriptsize
\begin{tabular}{l||c|cccc|c||c|cccc|c}\toprule
\multirow{3}{*}{Approach} &\multicolumn{6}{c||}{\juliet} &\multicolumn{6}{c}{\dataset} \\\cmidrule{2-13}
&Pred &\multicolumn{4}{c|}{Vul-CLS} &Vul-LOC &Pred &\multicolumn{4}{c|}{Vul-CLS} &Vul-LOC \\\cmidrule{2-13}
&Acc. &Acc. &Precision &Recall &F1 &Acc. &Acc. &Acc. &Precision &Recall &F1 &Acc. \\\midrule
\toolensem &\textbf{99.5\%} &\textbf{99.6\%} &\textbf{99.9\%} &\textbf{99.3\%} &\textbf{99.6\%} &\textbf{99.6\%} &\textbf{51.1\%} &58.9\% &\textbf{76.2\%} &48.1\% &\textbf{59.0\%} &\textbf{30.1\%} \\
\toolggnn &93.6\% &94.2\% &\textbf{99.9\%} &89.0\% &94.2\% &98.5\% &45.5\% &56.7\% &73.2\% &39.1\% &51.0\% &19.6\% \\
\tooltrans &99.3\% &\textbf{99.6\%} &\textbf{99.9\%} &\textbf{99.3\%} &\textbf{99.6\%} &99.1\% &47.2\% &\textbf{59.3\%} &70.5\% &\textbf{50.4\%} &58.8\% &29.3\% \\\midrule
Infer* &N/A &69.4\% &41.5\% &54.4\% &47.1\% &36.9\% &N/A &N/A &N/A &N/A &N/A &N/A \\
FlawFinder &N/A &58.3\% &36.6\% &51.0\% &42.6\% &8.6\% &N/A &48.7\% &53.3\% &6.0\% &10.7\% &6.7\% \\
RATS &N/A &59.3\% &36.5\% &46.4\% &40.9\% &N/A &N/A &45.8\% &47.9\% &16.4\% &24.2\% &N/A \\
\bottomrule
\end{tabular}
\vspace{-1mm}
\begin{flushleft}
\scriptsize 
*We do not compare with Infer on \dataset, since Zheng \etal~\cite{D2A} used Infer during data collection. Thus, it is unfair to compare with Infer on \dataset. 
\end{flushleft}
\vspace{-7mm}
\end{table*}
%\vspace{-2mm}
%\gail{why does one of the FSE reviewers say incomplete comparison to related work, particularly rule-based vulnerability detection?}

%In this section, we will present the evaluation of our proposed approach. We organize this section by asking research questions (RQs) and explain the concrete evaluation results accordingly.

We evaluate \tool with the following research questions:
%\rdc{list down all the RQs here}
\begin{itemize}[leftmargin=*]
    \item {\textbf{RQ1: Evaluating Classification \& Localization.} Can \tool locate vulnerabilities if they are not detected in advance?}
    \item {\textbf{RQ2: Evaluating Localization.} How does \tool perform on fine-grained vulnerability localization?}
    \item {\textbf{RQ3: Evaluating Ensemble Strategy.} Is the ensemble model an effective way to combine global and local context, compared with existing methods?}
    \item {\textbf{RQ4: Evaluating Fine-tuning Design.} What are the contributions of the \tools pre-training \& fine-tuning design?}
\end{itemize}

\vspace{-2mm}
\subsection{RQ1: Evaluating Classification \& Localization}
\label{subsec: rq2}

\noindent
\textbf{Motivation.}
First, we check how well \tool can locate vulnerable statements in a most realistic setting, where we do not know in advance whether the functions are vulnerable. This mimics a typical static analysis setting, where the static analyzer aims to find vulnerable lines. This experiment can be thought of as Classification \& Localization, where \tool will classify a function as vulnerable/non-vulnerable and then locate the statement within the vulnerable function.

In this first RQ, we focus on the comparison between \tool and the rule-based static analyzer baselines. 
%\gail{the rule-based nature of these static analyzers is not mentioned earlier}
We prioritize this discussion, since we aim at leveraging data-driven approaches to improve the quality of static analyzers in general, to further alleviate developers' debugging efforts. To comprehensively evaluate the effectiveness of our approach, we compare three variants of \tool with static analyzers: \toolensem, \toolggnn, \tooltrans. 

\noindent
\textbf{Methodology.} 
As mentioned in Section~\ref{subsec: prob_def}, to fit non-vulnerable functions into our framework, we add a dummy node with index of 0 to every sample regardless of its vulnerability. If the sample is non-vulnerable, then the ground-truth will be 0, the index of the dummy node; otherwise the ground-truth will still be the index of the vulnerable node.
Thus we successfully fit the vulnerability localization and the non-vulnerable function identification tasks into the same multi-class classification framework. We evaluate this task in:
%three settings:
%\vspace{-2mm}
\begin{enumerate}[leftmargin=*,noitemsep]
    \item 
    \textit{Multi-class prediction}: The most intuitive metric is class-prediction accuracy in the multi-class classification setting, and we refer to it as \emph{Prediction Accuracy} for further discussion. This setting is specific to our node-based classification and not applicable to static analyzers. 
    \item 
    \textit{Vulnerability Classification}: To evaluate \tools ability to detect vulnerabilities at function level, we define the predictions that point to the dummy node as non-vulnerable and otherwise as vulnerable. Note this setting is coarse-grained---if a vulnerable node has ground-truth $x (x > 0)$, but the model predicts non-dummy node $y (y > 0)$, where $x \neq y$, we still regard the prediction as correct in this setting. Thus, this setting basically evaluates whether there is a vulnerable node anywhere in the function body. 
    \item 
    \textit{Vulnerability Localization}: This is evaluated by top-1 accuracy, as we discussed in Section~\ref{sec:metric}, where we evaluate \tools ability to correctly predict vulnerable statements. We will not discuss this accuracy for RATS, since it tends to identify the root-cause location of a vulnerability (\eg line 16 of Listing~\ref{list:motivate}), yet our datasets annotate the location where the vulnerability occurs (\eg line 146 of Listing~\ref{list:motivate}). Thus, RATS has very bad accuracy (nearly 0\%) on our datasets. On the contrary, we confirmed that Infer and FlawFinder share our approach to location annotation. 
\end{enumerate}
\vspace{-1mm}

\noindent\textbf{Result: \toolbf significantly outperforms static analyzers, reducing false positives and false negatives.}
Table \ref{tab: rq1_full} shows the results when training the model on the combination of vulnerable and non-vulnerable functions. All variants under our \tool framework outperform static analyzers by a large margin, in three settings and on both datasets. The best-performing \toolensem achieves 99.6\% localization accuracy on \juliet and 30.1\% on \dataset, while the rule-based static analyzers, at best, only achieve 36.9\% and 6.7\%, respectively. 
Our data-driven approach also reported many fewer false positives (FP) and false negatives(FN) compared with the static analyzers. In the function-level vulnerability detection setting, the precision/recall/f1 of the \tool variations are significantly higher than all baseline static analyzers. The results empirically validate that \tool can help alleviate developers' concerns regarding the FP/FN issues of rule-based static analyzers. 

%\gail{Is there a technical explanation of why tool does so much better on synthetic than real-world?}

%\input{tables/rq2_results}

%\input{tables/rq1_prec_rec_f1}

%\vspace{-2mm}
%Knowing the better applicability and generalization of data-driven \tool than the rule-based tools, - don't know what this means
We further analyze \tools neural architecture module to figure out the best model for static analysis. Compared with the single GGNN and Transformer, the ensemble approach wins for \emph{Prediction Accuracy}, F1 and localization accuracy by a clear margin. The results, in general, reveal the effectiveness of the ensemble approach for combining two distinct architectures. Interestingly, we notice that Transformer beats the ensemble model by less than 1\% for classification accuracy. This is likely because the global view of the Transformer model enables it to have better performance on the more general vulnerability detection task, but this advantage is diluted by the local GGNN model when combining the two models. This leaves us an interesting question about how to improve the ensemble methodology when two models make contradictory predictions. We discuss a solution in Section~\ref{subsec: rq3}.

%\begin{result}
%\vspace{0.5mm}
%\textbf{Result.} \tool can beat rule-based static analyzers by a large margin on both vulnerability classification and localization tasks. The best \tool variant can correctly predict vulnerable statements with 99.6\% and 30.1\% accuracy on the synthetic and real-world data respectively. %\vspace{0.5mm}
%\end{result}

\vspace{-2mm}
\subsection{RQ2: Evaluating Localization}
%How does \toolbf perform on fine-grained vulnerability localization?}
\label{subsec: rq1}

\noindent
\textbf{Motivation.} After realizing the advantages of \tool, we are curious about the best-suited neural architecture for our primary goal, vulnerability localization. Thus, we further isolate this task by checking the efficiency of \tool variants only on the vulnerable samples. In other words, we check how efficiently \tool can identify a vulnerable statement given the function containing the statement is known to have a vulnerability. This mimics the scenario that by using some other off-the-shelf tools (\eg Devign~\cite{Zhou2019DevignEV}), we already know which functions are vulnerable. However, a C/C++ function may contain hundreds of lines of code, so it requires significant human effort to pinpoint the vulnerable location before attempting to fix it. To this end, for this RQ we only train the model on the vulnerable functions of our datasets. 

\noindent\textbf{Result-A: Ensemble model shows the best performance on localization.}
As shown in Table \ref{tab: rq1_result}, the ensemble approach wins against the single models, GGNN and Transformer, on both datasets for different metrics. For the \juliet dataset, 
%the ensemble approach is winning for both top-1 accuracy and average \emph{Prediction Distance}. Still, 
the benefits are not as pronounced since learning vulnerable patterns in this synthetic dataset is a relatively easy task for all models. For \dataset, the difference is more noticeable: by incorporating both global and local context learned by the two different architectures, \toolensem improves the top-1 accuracy of single \toolggnn by around 29.0\% and \tooltrans by 9.5\%. We also check the vulnerability types and find \tool is effective in locating integer overflow, buffer overrun, and null-pointer deferences, which aligns with the dominant types in the training data (Table~\ref{tab:data_distribution}). We also notice that the localization accuracy increases compared with those in Table \ref{tab: rq1_full} when isolating the localization task. The reason is, such an isolated training setting enables models to focus on learning the vulnerable triggering locations and thus decrease the overall difficulty of learning to predict the vulnerable nodes. This implies that if a fairly good function vulnerability detector is available, training a localizer with only vulnerable samples will be a better choice.

\begin{table}[h]\centering
\vspace{-3mm}
\caption{\small Test performance of vulnerability localization under \tool framework. 
Top-k represents the localization accuracy for top-k predictions. Distance represents the average \emph{Prediction Distance} between the prediction and the ground-truth.}
\vspace{-1mm}
\label{tab: rq1_result}
\footnotesize
\begin{tabular}{lrr|rrrr}\toprule
\multirow{2}{*}{\tool} &\multicolumn{2}{c|}{Juliet} &\multicolumn{3}{c}{D2A} \\\cmidrule(lr){2-3}\cmidrule(lr){4-6}
&Top-1 &Distance &Top-1 &Top-3 &Distance \\\midrule
Ensemble &\textbf{99.6\%} &\textbf{0.04} &\textbf{43.6\%} &\textbf{63.9\%} &\textbf{7.0} \\
GGNN &98.1\% &0.10 &33.8\% &54.9\% &9.7 \\
Transformer &99.4\% &0.06 &39.8\% &62.4\% &8.0 \\
\bottomrule
\end{tabular}
\vspace{-3mm}
\end{table}

Further, \tool gives us a decent \emph{Prediction Distance}, indicating that the vulnerabilities can be located within a 7-line window around the ground-truth vulnerable statements. The ensemble setting not only beats the baselines but, overall, it is a respectable scope-reducer for debugging as the average length of a \dataset function is 80.3 lines and a significant fraction of \dataset functions have hundreds of source lines.
% it would be better if we could say what that fraction is

%\gail{one reviewer makes a big deal about the set of baselines only including open-source static analyzers rather than commercial analyzers. But the response says Infer *is* a commercial analyzer, shouldn't this be mentioned in the body of the paper? This reviewer keeps pointing out things like "FlawFinder is a very weak adversary".  why is this?}

\noindent
\textbf{Result-B: Transformer learns the global context, while GGNN captures the local context.} 
We have empirically demonstrated that global and local context together can improve vulnerability localization. We further investigate the performance of \toolggnn and \tooltrans on the 133 (most recent) vulnerabilities in \dataset -{\sc test} %(see Figure \ref{fig:venn_diagram})
---\textit{GGNN and Transformer are manifesting complementary capacities to localize the vulnerability}. Due to the repetition and simplicity of certain vulnerable patterns, two models locate 37 vulnerabilities in common; besides these, GGNN can further locate 8 individual vulnerabilities and Transformer alone can locate 16. 

%\vspace{-1mm}

We inspect the concrete samples that one model correctly predicts but the other fails.  We compute the average function length of the model-specific correct predictions. As expected, Transformer's correct predictions have a larger function length than for GGNN: GGNN's correct predictions include functions with, on average, 27.75 lines and 138.0 graph nodes; for Transformer's correct predictions, however, the functions include 48.38 lines and 288.6 graph nodes, on average, which are 74.3\% longer in lines and 109.1\% larger in nodes than GGNN. We further conduct statistical test to make sure that the longer average value is not caused by extremely large outliers.
% but the previous section says there is one function with 1269 lines! it's not an outlier?
This result supports our intuition that Transformer has a better \emph{global} view and is more effective in processing larger code graphs, whereas GGNN focuses more on the \emph{local} contexts of code snippets. We also show two concrete vulnerabilities in \dataset-{\sc Test}: Listing~\ref{list:transonly}\&\ref{list:ggnnonly}. Both can be located by the ensemble approach, but the single model fails on one of them. Listing \ref{list:transonly} has a larger size and the global contexts of \texttt{out}, \texttt{mtmp} and \texttt{outlen} are necessary to identify buffer overflow. In contrast, Listing \ref{list:ggnnonly} is much smaller and the variable dependencies are mostly in the surrounding lines.

%We also notice that if we directly consider the union of Transformer and GGNN, it contains 61 correct predictions, out of 133 test samples, while our ensemble model correctly predicts 58. This decrease is mainly due to, the wrongly predicted single model gives very high confidence on some samples, while the correct model gives lower confidence on these samples. In such cases, the wrong prediction dominates during the ensemble, so that the final prediction is also wrong. In practice, it is infeasible to do the ``direct union" as we no more have the ground-truth for unseen data, to tell which model is making a correct prediction. Thus, two models might not be enough in reality since they sometimes make contradictory predictions on the unseen data. We provide a solution for this in Section \ref{subsec: rq3}

%\vspace{-1mm}

%\begin{result}
%\vspace{0.5mm}
%\textbf{Result.} Combining global and local context, \tool can successfully locate real-world vulnerabilities with 43.6\% top-1 and 63.9\% top-3 accuracies. 
%\vspace{0.5mm}
%\end{result}

%\gail{did you try some weighting besides averaging? the response says something about two of each kind, but I cannot tell if they're still all averaged.  The paper could claim something like "future work" will empirically evaluate different weighting schemes and mechanisms that forces the specialisation of the sub-components on different parts of the problem space - mentioned by the second reviewer.}

\vspace{-2mm}
\begin{lstlisting}[style=customc, belowcaptionskip=0pt, caption=\small Buffer Overrun vulnerability from \dataset. \toolensem and \tooltrans can localize this vulnerability at line 42 while \toolggnn fails., label={list:transonly}, escapechar=\%]
// project: openssl (commit sha: 0211740)
// file: crypto/ec/ecdh_kdf.c
1  int ecdh_KDF_X9_63(...) 
2  {...
18 for (i = 1;; i++) {
19   unsigned char mtmp[EVP_MAX_MD_SIZE];
     ...
32   if (outlen >= mdlen) {
33   if (!EVP_DigestFinal(mctx, out, NULL))
34     goto err;
35   outlen -= mdlen;
36   if (outlen == 0)
37     break
     ...
42   @memcpy(out, mtmp, outlen);@ ...
69 }
\end{lstlisting}
\vspace{-3mm}
\begin{lstlisting}[style=customc, belowcaptionskip=0pt, caption= \small Integer Overflow vulnerability from \dataset. \toolensem and \toolggnn can localize this vulnerability at line 19 while \tooltrans fails., label={list:ggnnonly}, escapechar=\%]
// project: ffmpeg
// sha: 4bd869e
// libavcodec/aacdec.c
1  static int read_audio_mux_element(...) 
2  {
3    int err;
4    uint8_t use_same_mux = get_bits(gb, 1);
     ...
14   if (latmctx->audio_mux_version_A == 0) {
15     int mux_slot_length_bytes = read_payload_length_info(latmctx, gb);
     ...
19     @else if (mux_slot_length_bytes * 8 + 256 < get_bits_left(gb)) {@ ...
25   return 0;
26 }
    
\end{lstlisting}

\vspace{-3mm}

\subsection{RQ3: Evaluating Ensemble Strategy}
\label{subsec: rq3}

\noindent
\textbf{Motivation.} We have shown the effectiveness of the ensemble model for comprehensively understanding the \emph{global} and \emph{local} contexts, compared with the single model. However, the ensemble approach is not the only way to achieve this aggregation for code modeling tasks, as discussed in Section~\ref{sec:baselines}.  
We incorporate the alternate aggregation strategies to better evaluate the ensemble approach. We train the models in two settings: first, we only include the vulnerable samples to reveal the localization performance (Vul-only), and then we expand the data to contain both vulnerable and non-vulnerable functions (Hybrid). To evaluate performance, we use the \emph{Prediction Accuracy} defined in Section \ref{subsec: rq2}.

\begin{table}[!htp]\centering
\vspace{-2mm}
\caption{\small Comparison of \emph{Prediction Accuracy} between our ensemble method and the existing models. 
}\label{tab: rq3_results}
\scriptsize

\begin{tabular}{lrrrrr}\toprule
\multirow{2}{*}{Model} &\multicolumn{2}{c}{\juliet} &\multicolumn{2}{c}{\dataset} \\\cmidrule(lr){2-3}\cmidrule(lr){4-5}
&Vul-only &Hybrid &Vul-only &Hybrid \\\midrule
{\sc Velvet-ensemble} &\textbf{99.6\%} &\textbf{99.5\%} &\textbf{43.6\%} &\textbf{51.1\%} \\
{\sc Velvet-ggnn} &98.1\% &93.6\% &33.8\% &45.5\% \\
{\sc Velvet-transformer} &99.4\% &99.3\% &39.8\% &47.2\% \\
{\sc Velvet-TransSandwich} &99.2\% &99.0\% &41.4\% &49.8\% \\
{\sc Velvet-great} &99.1\% &98.8\% &36.8\% &41.6\% \\
{\sc Velvet-hoppity} &98.8\% &98.1\% &35.3\% &32.5\% \\
\bottomrule
\end{tabular}
\vspace{-2mm}
\end{table}

\smallskip
\noindent
\textbf{Result-A: \toolbf's ensemble strategy outperforms existing models that aggregate global and local contexts. }
As shown in Table \ref{tab: rq3_results}, the ensemble approach manifests a more powerful learning capacity compared with the existing models we studied. \toolensem beats all other competitors for both vulnerability-only and hybrid settings on both datasets. Specifically, on the real-world \dataset, \toolensem illustrates the generalization of the ensemble approach to understand the complicated real-world vulnerable patterns, compared with the three re-implemented existing works. This result shows empirically that the ensemble model is a more direct and effective way to combine \emph{global} and \emph{local} knowledge than stacking distinct models, since the stack of models will still share learned insights  during training, which may prevent them from learning more diversity. However, compared with the single GGNN and Transformer, the Transformer-sandwich model still reports overall better results with a clear margin on \dataset. This also suggests the significance of incorporating distinct model designs to capture varied aspects of code.

\noindent\textbf{Result-B: Ensemble of more models alleviates the disagreement among models and further improves performance.} Our main goal is to showcase the advantages of ensemble learning with local and global information, so we initially use one global model and one local model to conceptually illustrate the effectiveness. However, as mentioned in Section \ref{subsec: rq2}, Transformer and GGNN sometimes make contradictory predictions and the dominant model is not always correct. Consequently, such disagreements harm the ensemble's performance. We provide a solution to alleviate this concern: adding more models of the same architecture to the ensemble. As a proof-of-concept, we enlarge \toolensem to contain \emph{two} Transformer and \emph{two} GGNN models by varying the initialization seeds, and we compare the 4-model variant of \tool with the 2-model one in both Vul-only and Hybrid settings on \dataset. Table~\ref{tab: more_ensemble_model} shows 4-model \tool is a clear winner and it can also beat all baselines in previous RQs. The results further show the effectiveness of the ensemble approach and indicate \tools potential practical deployment. 

\begin{table}[h]\centering
\vspace{-2mm}
\caption{\small Performance of adding one more GGNN and one more Transformer model to the \toolensem. The newly added models are randomly initialized and have exactly the same configuration with the original ones.}\label{tab: more_ensemble_model}
\scriptsize
\begin{tabular}{cccc|rrrr}\toprule
\multirow{3}{*}{\tool}
&\multicolumn{3}{c}{\dataset Vul-Only} &\multicolumn{3}{c}{\dataset Hybrid} \\\cmidrule{2-7}
&\multicolumn{2}{c}{Vul-LOC Acc.} & &Pred &Vul-CLS &Vul-LOC \\\cmidrule{2-3}
&Top-1 &Top-3 &Distance &Acc. &Acc. &Acc. \\\midrule
2-model &43.6\% &63.9\% &7.0 &51.1\% &58.9\% &30.1\% \\
4-model &\textbf{45.9\%} &\textbf{68.4\%} &\textbf{6.5} &\textbf{52.4\%} &\textbf{60.6\%} &\textbf{31.6\%} \\
\bottomrule
\end{tabular}
\vspace{-2mm}
\end{table}

\vspace{-2mm}
\subsection{RQ4: Evaluating Fine-Tuning Design}
\label{subsec: rq4}

%\noindent\textbf{Advantage of Ensemble Learning.}

%\noindent\textbf{Fine-tuning.}
\noindent
\textbf{Motivation.} In Section \ref{sec:design}, we introduced the setting of applying fine-tuning based on the pre-trained \juliet models to real-world data, to try to address the data inadequacy problem. In this RQ, we evaluate the rationality and advantages of this design. Theoretically, we expect the pre-trained \juliet models to already understand a portion of the real-world vulnerable patterns, since the synthetic data is designed to imitate the practical scenario, even as it struggles with the complexity of the real-world distribution. To understand whether this is the case, we evaluate the well-trained \juliet models of \toolensem directly on \dataset to see how much knowledge has been directly transferred from the synthetic data. Again, we use the two same settings, Vul-only and Hybrid, as in Section \ref{subsec: rq3} for this evaluation.
%\vspace{-2mm}
\begin{table}[!htp]\centering
\caption{\small Results of \toolensem before and after the fine-tuning on two datasets, with two settings. The ``Before" row means we directly evaluate the well-trained \juliet model on \dataset. The ``After" row means the model's performance after the sufficient fine-tuning on real-world data.}
\label{tab: rq4_result_a}
\scriptsize
\begin{tabular}{lrrrrrr}\toprule
\multirow{3}{*}{Fine-Tuning} &\multicolumn{2}{c}{Vul-only} &\multicolumn{3}{c}{Hybrid} \\\cmidrule(lr){2-3}\cmidrule(lr){4-6}
&Vul-LOC &Distance &Pred &Vul-CLS &Vul-LOC \\
&ACC & &ACC &ACC &ACC \\\midrule
Before &12.0\% &14.8 &16.5\% &56.3\% &3.8\% \\
After &43.6\% &7.0 &51.1\% &58.9\% &30.1\% \\
\bottomrule
\end{tabular}
\vspace{-5mm}
\end{table}
%\vspace{-2mm}

%\smallskip
\noindent
\textbf{Result: Pre-training on synthetic data and fine-tuning on real-world data effectively mitigates the real-world vulnerability samples inadequacy problem.} 
% too strong to say it solves the problem
As shown in the "Before" row of Table~\ref{tab: rq4_result_a}, the pre-trained \juliet model is already able to correctly localize 12\% of \dataset test samples, with an average prediction distance of 14.8 in the Vul-only setting, and has 16.5\% \emph{Prediction Accuracy} for the Hybrid setting. This result shows that a significant portion of real-world vulnerable patterns are already well-understood by the pre-trained \juliet models, providing a great start for fine-tuning. Also, with its knowledge of tens of thousands of synthetic samples, the model is less likely to be overfitted to the relatively small \dataset dataset. As a comparison, we also show the performance after the fine-tuning (\ie``After" row in Table~\ref{tab: rq4_result_a}). We can see that the fine-tuning significantly improves the performance on the real-world dataset, and even with just 2.5k vulnerable samples and 2.5k non-vulnerable samples, the model can show a generalized result in both Vul-only and Hybrid settings. The results also provide solid evidence to the rationality of our mitigation of the data inadequacy, which hopefully can help researchers move forward without being too worried about small real-world dataset sizes for data-driven approaches.

\section{Related Work}
\label{sec: related}

\noindent\textbf{Fault Localization.} Locating buggy statements with the available test cases has been well-studied for decades~\cite{Abreu2006Ochiai, Abreu2007accuracy, Lee2011Spectrum, Jones2005Tarantula, moon2014mutate, Papadakis2012mutate, Zhang2013mutate, li2019deepfl, zhang2019cnnfl, lou2020can}.
Spectrum-based (SBFL) and mutation-based (MBFL) fault localization are two well-known approaches. SBFL takes each statement's coverage information and its test case results as input, calculates a suspiciousness score, and ranks the statements' scores to indicate the most buggy one(s). MBFL mutates the code by pre-defined rules to evaluate each statement's actual effects on the pass/fail outcomes of test cases.  \tool is not directly comparable, since we focus on source code without test cases and coverage reports.

\noindent\textbf{DL-based Vulnerability Detection.} To automatically learn vulnerable features and patterns directly from source code, recent work~\cite{li2018vuldeepecker, li2018sysevr, suneja2020learning, russell2018automated, li2021vuln, Zhou2019DevignEV, chakraborty2020deep, xiao2021deepwukong} applies a wide variety of deep-learning models.
However, most of this work predicts function-level vulnerability, even for functions containing up to thousands of lines. Even when developers know a function is vulnerable, 
% unsafe, \gail{unsafe seems wrong word}
they must spend 
%plenty of 
time to locate the specific statements to edit. 
%problematic statements and repair them. 
Wang~\etal~\cite{wang2020ginn} propose GINN, an advanced GGNN model, and show
%reveal 
its ability to localize null pointer dereferences, among several GINN applications. This work seems closest to ours
%They are the most related work to us. 
but their research direction is orthogonal:
%with ours: 
we regard graph-based models in general,
%as a whole kind, 
ignoring the subtle differences among variants; we instead study the complimentary effects between graph-based and sequence-based models, and leverage %such 
this distinction to better localize vulnerabilities. So for our proof of concept, we just pick the popular GGNN architecture as baseline.
\section{Threats to Validity}
\label{sec:threats}

%gail{these threats to validity need to be addressed here if they're not elsewhere in the paper:%Vulnerability detection is based on balanced dataset, which is unrealistic and doesn't show how precise the approach is in practice - why wasn't tool applied to some real-world project(s) with a couple known vulnerabilities, for a more realistic unbalanced setting?
%Real-world dataset is based on warnings by Infer, i.e., the model presumably can only be as good Infer.  as the reviewer notes, why not just use Infer?
%wrt ground truth: was there any manual inspection of a random sample?}

\noindent
\textbf{Cross Validation.} We tried to imitate a pragmatic real-world scenario, where we train the model on the past vulnerabilities and test on the latest samples. However, due to the small size of the \dataset-{\sc test} split, the model's reported results may not be generalizable. To minimize this threat, we did ten-fold cross-validation for all baseline models in the setting of Section \ref{subsec: rq1}. Table \ref{tab: threats} reveals the same trend that \toolensem is winning by a clear margin on both metrics.
\vspace{-2mm}
\begin{table}[!htp]\centering
\caption{\small Ten-fold cross-validation for Section \ref{subsec: rq1}}\label{tab: threats}
\scriptsize
\begin{tabular}{lrrr}\toprule
&Vul-LOC Acc &Distance \\\midrule
\toolensem &\textbf{39.9}\% &\textbf{8.6} \\
\toolggnn &34.4\% &9.8 \\
\tooltrans &37.9\% &9.1 \\
{\sc Velvet-TransSand} &39.3\% &9.5 \\
{\sc Velvet-great} &37.9\% &8.7 \\
{\sc Velvet-hoppity} &35.5\% &10.1 \\
\bottomrule
\end{tabular}
\vspace{-2mm}
\end{table}

%\gail{the second reviewer notes the 10-fold results here, but only for the problem of localisation distance, but response implies there's more. Maybe due to lack of space... Since the reviewer missed that table 5 includes vul-loc, put that kind of thing in the things like column/row headers or captions, not just in prose.}

%\gail{the response to reviewers says that intra-project performance was worse than cross-project, which is counterintuitive, why is this?}

\smallskip
\noindent
\textbf{Cross-project.} \tool focuses on intra-project vulnerable patterns, since we expect the model to learn from the project's history and apply its knowledge to the project's new commits. To minimize the threats brought by such settings, we further study the model's cross-project performance under the Section~\ref{subsec: rq1} setting. We pick the \dataset \texttt{Apache-HTTPD} project for evaluation and the other projects for training. The results show \tool can achieve 41.8\% Vul-LOC accuracy for top-1 and 60.0\% for top-5, dropping a bit compared with the intra-project setting but still very promising.

\smallskip
\noindent
\textbf{Data Generalizability.} Our results might be biased by the underlying data collection process, conducted by Zheng \etal~\cite{D2A}.  However, collecting vulnerability data at statement granularity is hard. We have not found alternative real-world data with such fine-granular information to fully evaluate generalizability. To minimize this threat, we also evaluate \tool on a synthetic dataset and show it performs better than static analysis alternatives.

%\gail{this does not sound good, a synthetic dataset is not magically unbiased.  maybe say something about the 'unbiased' properties of that synthetic dataset or that its widely used.}

%\rdc{Talk about train the model end-to-end, with classification model}

%\vspace{-2mm}
\section{Conclusion}
\label{sec:conclusion}
We introduced \tool, an ensemble model to efficiently capture local and global code context and understand vulnerable patterns at the individual statement level. Our evaluation showed that \tool is effective for two vulnerability localization tasks on both synthetic and real-world data. Our designed workflow of pre-training on synthetic data and then fine-tuning on real-world data provides a practical solution to the real-world dataset scarcity problem. 

%\gail{the paper needs a discussion somewhere about generalizing this approach to fault localization, e.g., in the context of APR, not specifically vulnerable statements. if there's anything special about vulnerable statements that requires a different appraoch than more general fault localization, the paper needs to say so, but I'd guess the only issue is you have datasets for vulnerability detection and did not consider datasets for fault localization like dbgbench}

\section*{Acknowledgment}
This work is supported in part by NSF grants CCF-2107405, CCF-1845893, CCF-1815494, IIS-2040961, DARPA/NIWC-Pacific N66001-21-C-4018, and IBM. Any opinions, findings, conclusions, or recommendations expressed herein are those of the authors and do not necessarily reflect those of the US Government, NSF, DARPA, or IBM.

%\newpage
\bibliographystyle{ieeetr}
\bibliography{main, bray}

\end{document}